\documentclass{aa}
\usepackage{graphics}
 
\newcommand{\HI}{\mbox{H{\sc i}}}

\begin{document}      

   \title{A dynamical model for the heavily ram pressure stripped Virgo spiral galaxy NGC~4522}

   \author{B.~Vollmer\inst{1}, M.~Soida\inst{2}, K.~Otmianowska-Mazur\inst{2}, 
     J.D.P.~Kenney\inst{3}, J.H.~van~Gorkom\inst{4}, \& R.~Beck\inst{5}}

   \offprints{B.~Vollmer, e-mail: bvollmer@mpifr-bonn.mpg.de}

   \institute{CDS, Observatoire astronomique de Strasbourg, 11, rue de l'universit\'e,
	      67000 Strasbourg, France \and
	      Astronomical Observatory, Jagiellonian University,
	      Krak\'ow, Poland \and
	      Yale University Astronomy Department, P.O. Box 208101, New Haven, CT 06520-8101, USA \and
	      Department of Astronomy, Columbia University, 538 West 120th Street, New York, 
	      NY 10027, USA \and
	      Max-Planck-Insitut f\"{u}r Radioastronomie, Auf dem H\"{u}gel 69, 53121 Bonn, Germany
              }

   \date{Received / Accepted}

   \authorrunning{Vollmer et al.}
   \titlerunning{A dynamical model for NGC~4522}

\abstract
{Evolution of spiral galaxies in the Virgo cluster.}
{We determined the parameters of the interaction between the interstellar medium of 
NGC~4522 and the intracluster medium of the Virgo cluster.}
{A dynamical model including ram pressure stripping is applied to the strongly H{\sc i} deficient
Virgo spiral galaxy NGC~4522. A carefully chosen model snapshot is compared with existing 
VLA H{\sc i} observations.}
{The model successfully reproduces the large-scale gas distribution and the velocity field. However it
fails to reproduce the large observed H{\sc i} linewidths in the extraplanar component, for
which we give possible explanations. In a second step, we solve the induction equation on the
velocity fields of the dynamical model and calculate the large scale magnetic field. Assuming a
Gaussian distribution of relativistic electrons we obtain the distribution of polarized radio 
continuum emission which is also compared with our VLA observations at 6~cm. The observed maximum
of the polarized radio continuum emission is successfully reproduced. Our model suggests that
the ram pressure maximum occurred only $\sim 50$~Myr ago.}
{Since NGC~4522 is
located far away from the cluster center ($\sim 1$~Mpc) where the intracluster medium density
is too low to cause the observed stripping if the intracluster medium is static and smooth, 
two scenarios are envisaged: (i) the galaxy moves
very rapidly within the intracluster medium and is not even bound to the cluster;
in this case the galaxy has just passed the region of highest intracluster medium density;  
(ii) the intracluster medium is not static but moving due to the infall of the M49
group of galaxies. In this case the galaxy has just passed the
region of highest intracluster medium velocity.
This study shows the strength of combining high resolution H{\sc i} and polarized radio continuum
emission with detailed numerical modeling of the evolution of the gas and the large-scale
magnetic field.}


\maketitle

\section{Introduction \label{sec:intro}}

The spiral galaxy NGC~4522 is one of the best examples for ongoing ram pressure 
stripping due to the galaxy's rapid motion within the hot and tenuous intracluster gas (ICM) of the 
Virgo cluster. H{\sc i} and H$\alpha$ observations (Kenney et al. 2004, 
Kenney \& Koopmann 1999) showed a
heavily truncated gas disk at a radius of 3~kpc, which is $\sim 40$\% of the optical radius,
and a significant amount of extraplanar gas to the west 
of the galactic disk. The one-sided extraplanar atomic gas distribution shows high column
densities, comparable to those of the adjectant galactic disk. 
Such high extraplanar gas column densities are unusual among Virgo 
spiral galaxies (Cayatte et al. 1990), suggesting that NGC~4522 is in a 
short-lived phase of evolution.

Since the stellar disk is symmetric and undisturbed (Kenney \& Koopmann 1999), 
a tidal interaction is excluded as the origin of the peculiar 
gas distribution of NGC~4522. A scenario where NGC~4522 is moving roughly to the east 
and experiences strong ram pressure 
can account for the truncated gas disk and the western extraplanar gas. 
A scenario where the ram pressure peak occurred a few 100~Myr ago 
(stripping by the intracluster medium around M87; Vollmer et al. 2000) is excluded,
because the H{\sc i} column density distribution and velocity field are inconsistent
with fall back of stripped gas which should occur a few 100~Myr after peak 
ram pressure, which is needed to create extraplanar gas at these late times. 
Further evidence for the peak ram pressure
scenario comes from polarized radio continuum observations (Vollmer et al. 2004).
The 6~cm polarized emission is located at the eastern edge of the galactic disk, opposite to the 
western extraplanar gas. This ridge of polarized radio continuum emission is most likely
due to ram pressure compression of the interstellar medium (ISM) and its magnetic field. 
In addition, the degree of polarization decreases from the east to the west and the flattest 
spectral index between 20~cm and 6~cm coincides with the peak of the 6~cm polarized emission. 
These findings are also consistent with a scenario where ram pressure is close to its maximum.

This scenario has one important caveat: NGC~4522 is located at a projected distance of $\sim 1$~Mpc
from the center of the Virgo cluster (M87). Assuming a static smooth ICM and standard values
for the ICM density and the galaxy velocity, the ram pressure at that location
seems to be too low by an order of magnitude to produce the observed truncation of the gas disk. 
Therefore, Kenney et al. (2004) and Vollmer et al. (2004) claim that either
(i) NGC~4522 has a velocity of $\sim 4000$~km\,s$^{-1}$ with respect to the Virgo
cluster mean or (ii) that the ICM density is locally enhanced or 
(iii) that the intracluster medium is moving or (iv) a combination of (ii) and (iii).
Based on existing X-ray observations Kenney et al. (2004) claim that the
latter scenario is more likely, given that
NGC~4522 is located between M87 and the bright elliptical galaxy M49 which 
is the center of a galaxy subcluster falling into the Virgo cluster from behind
(Irwin \& Sarazin et al. 1996 and Biller et al. 2004).  In addition, the X-ray hotspot detected by
Shibata et al. (2001) represents the sign of an interaction between the
Virgo intracluster medium and that of the M49 subcluster. The intracluster medium of the
M49 subcluster has thus a velocity whose radial component is opposite to the
radial velocity of NGC~4522. This moving intracluster medium together with a modest 
local density enhancement of the intracluster medium by a factor of 2--4
can account for the observed stripping radius.

In this article we present a ram pressure stripping model for NGC~4522 in order to
assess if a simulation using a time dependent ram pressure 
can account for the observed gas distribution, velocity field, and polarized radio continuum
emission distribution. Our aim is to constrain the parameters of the event. 
Since the observations of Kenney et al. (2004) and Vollmer et al. (2004)
showed that ram pressure is ongoing, we only consider model snapshots close to peak ram pressure. 

In Sect.~\ref{sec:model} we describe the dynamical model. The choice of the best fit
model is justified and its time evolution and the final snapshot for the comparison 
with observations is presented in Sect.~\ref{sec:bestfit}. 
The comparison with VLA
H{\sc i} data is done using moment maps and position-velocity diagrams (Sect.~\ref{sec:comparison}).
We solved the induction equation for the velocity field of our simulation to obtain
the evolution of the large scale magnetic field and the distribution of the 
polarized radio continuum emission which is compared with observations in Sect.~\ref{sec:mhd}.
We discuss our results in Sect.~\ref{sec:discussion} and give our conclusions in 
Sect.~\ref{sec:conclusions}.

\section{Dynamical model \label{sec:model}}

We use the N-body code described in Vollmer et al. (2001) which consists of 
two components: a non-collisional component
that simulates the stellar bulge/disk and the dark halo, and a
collisional component that simulates the ISM.

The non--collisional component consists of 49\,125 particles, which simulate
the galactic halo, bulge, and disk.
The characteristics of the different galactic components are shown in
Table~\ref{tab:param}.
\begin{table}
      \caption{Total mass, number of particles $N$, particle mass $M$, and smoothing
        length $l$ for the different galactic components.}
         \label{tab:param}
      \[
         \begin{array}{lllll}
           \hline
           \noalign{\smallskip}
           {\rm component} & M_{\rm tot}\ ({\rm M}$$_{\odot}$$)& N & M\ ({\rm M}$$_{\odot}$$) & l\ ({\rm pc}) \\
           \hline
           {\rm halo} & 4.8\,10$$^{10}$$ & 32768 & $$1.48\,10^{6}$$ & 1200 \\
           {\rm bulge} & 1.7\,10$$^{9}$$ & 16384 & $$1.0\,10^{5}$$ & 180 \\
           {\rm disk} & 8.3\,10$$^{9}$$ & 32768 & $$2.5\,10^{5}$$ & 240 \\
           \noalign{\smallskip}
        \hline
        \end{array}
      \]
\end{table}
The resulting rotation velocity is $\sim$100~km\,s$^{-1}$ and the rotation curve
becomes flat at a radius of about 4~kpc. 

We have adopted a model where the ISM is simulated as a collisional component,
i.e. as discrete particles which possess a mass and a radius and which
can have inelastic collisions (sticky particles).
Since the ISM is a turbulent and fractal medium (see e.g. Elmegreen \& Falgarone 1996),
it is neither continuous nor discrete. The volume filling factor of the warm and cold phases
is smaller than one. The warm neutral and ionized gas fill about $30-50\%$ of the volume,
whereas cold neutral gas has a volume filling factor smaller than 10\% (Boulares \& Cox 1990). 
It is not clear how this fraction changes, when an external 
pressure is applied. In contrast to smoothed particles hydrodynamics (SPH), which is a 
quasi continuous approach and where the particles cannot penetrate each other, our approach 
allows a finite penetration length, which is given by the mass-radius relation of the particles.
Both methods have their advantages and their limits.
The advantage of our approach is that ram pressure can be included easily as an additional
acceleration on particles that are not protected by other particles (see Vollmer et al. 2001).
In this way we avoid the problem of treating the huge density contrast between the 
ICM ($n \sim 10^{-4}$~cm$^{-3}$) and the ISM ($n > 1$~cm$^{-3}$) of the galaxy.

The 20\,000 particles of the collisional component represent gas cloud complexes which are 
evolving in the gravitational potential of the galaxy.
The total assumed gas mass is $M_{\rm gas}^{\rm tot}=1.3\,10^{9}$~M$_{\odot}$,
which corresponds to the total neutral gas mass before stripping, i.e.
to an \HI\ deficiency of 0, which is defined as the logarithm of the ratio between
the \HI\ content of a field galaxy of same morphological type and diameter
and the observed \HI\ mass.
To each particle a radius is attributed depending on its mass. 
During the disk evolution the particles can have inelastic collisions, 
the outcome of which (coalescence, mass exchange, or fragmentation) 
is simplified following Wiegel (1994). 
This results in an effective gas viscosity in the disk. 

As the galaxy moves through the ICM, its clouds are accelerated by
ram pressure. Within the galaxy's inertial system its clouds
are exposed to a wind coming from a direction opposite to that of the galaxy's 
motion through the ICM. 
The temporal ram pressure profile has the form of a Lorentzian,
which is realistic for galaxies on highly eccentric orbits within the
Virgo cluster (Vollmer et al. 2001).
The effect of ram pressure on the clouds is simulated by an additional
force on the clouds in the wind direction. Only clouds which
are not protected by other clouds against the wind are affected.

The particle trajectories are integrated using an adaptive timestep for
each particle. This method is described in Springel et al. (2001).
The following criterion for an individual timestep is applied:
\begin{equation}
\Delta t_{\rm i} = \frac{20~{\rm km\,s}^{-1}}{a_{\rm i}}\ ,
\end{equation}
where $a_{i}$ is the acceleration of the particle i.
The minimum value of $t_{\rm i}$ defines the global timestep used 
for the Burlisch--Stoer integrator that integrates the collisional
component.

\section{Search for the best fit model \label{sec:bestfit}}

In this section we will constrain the parameters of the ram pressure stripping event
which are (i) the peak ram pressure, (ii) the temporal ram pressure profile,
(iii) time since peak ram pressure, (iv) the inclination angle between the galaxy's 
disk and the intracluster medium wind direction, and (v) the azimuthal viewing angle for 
the observed inclination and position angles.
These parameters are related to the observed quantities which are (1) the position angle,
(2) the inclination angle of the galactic disk, (3) the line-of-sight velocity
of the galaxy with respect to the cluster mean, and (4) the projected ICM wind direction.
The position angle and inclination of NGC~4522 define a plane in three dimensional space.
The model galaxy can then be rotated within this plane by the azimuthal viewing angle (see below). 
The three dimensional model wind direction, the line-of-sight velocity of the galaxy, and
the projected ICM wind direction are thus functions of the azimuthal viewing angle.

In the case of a smooth static ICM ram pressure is proportional to the ICM density
$\rho_{\rm ICM}$ and the square of the galaxy velocity with respect to the Virgo cluster 
$\vec{v_{\rm gal}}$. If the ICM is moving with respect to the cluster mean velocity
the expression for ram pressure yields:
\begin{equation}
p_{\rm ram}=\rho_{\rm ICM} (\vec{v_{\rm gal}}-\vec{v_{\rm ICM}})^{2}\ ,
\label{eq:icmwind}
\end{equation}
where $\vec{v_{\rm ICM}}$ is the galaxy's velocity vector with respect to the Virgo cluster, 
and $\vec{v_{\rm gal}}$ is the galaxy velocity with respect to the Virgo cluster.

As noted in Kenney et al. (2004) the orbit of NGC~4522 is not easy to understand.
Its high radial velocity ($\sim 1300$~km\,s$^{-1}$) with respect to the cluster 
mean velocity excludes a classical radial orbit. 
At its projected distance from the cluster center (3.3$^{\circ} \sim 1$~Mpc)
the density of the intracluster medium is an order of magnitude too low to strip the galaxy's
ISM up to a galactic radius of 3~kpc if one assumes a velocity of the galaxy
of $1500$~km\,s$^{-1}$ within the cluster (Kenney et al. 2004). Kenney et al. (2004) and
Vollmer et al. (2004) propose two possible scenarios
enhancing the ram pressure by a factor of 10: (i) the ICM is moving opposite
to the trajectory of NGC~4522 due to the infall of the M49 group into the
Virgo cluster and (ii) NGC~4522 is not bound to the cluster
and crosses it with a velocity of $\sim 4000$~km\,s$^{-1}$ only once. 
In both cases we expect a strongly peaked temporal ram pressure profile similar to that
for radial orbits:
\begin{equation}
p_{\rm ram}=p_{\rm max} \frac{t_{\rm HW}^{2}}{t^{2}+t_{\rm HW}^{2}}\ ,
\label{eq:rps}
\end{equation}
where $t_{\rm HW}$ is the width of the profile (Vollmer et al. 2001). 
Since the observations of Kenney et al. (2004) and Vollmer et al. (2004)
suggest that ram pressure is close to maximum, we only consider model snapshots close
to peak ram pressure $p_{\rm max}$, which has to be at least 1000~cm$^{-3}$(km\,s$^{-1}$)$^{2}$ 
to strip gas with a column density of 10~M$_{\odot}$pc$^{-2}$ at the observed
stripping radius of 3~kpc (Kenney et al. 2004). We do not consider higher ram
pressure maxima, because this value is already a factor 10 above the
estimate based on a static smooth intracluster medium and a velocity of
$\sim 1000$~km\,s$^{-1}$. The Gaussian half width $t_{\rm HW}$
is not critical as long as it is not too large. Since the velocity of NGC~4522
with respect to the intracluster medium (flowing or not) is high, a value
between 50 and 100~Myr is reasonable.
Thus, we set $p_{\rm max}$=2000~cm$^{-3}$(km\,s$^{-1}$)$^{2}$ and $t_{\rm HW}$=80~Myr.
We define $t=0$~Myr as the time when ram pressure is maximum.

The ram pressure efficiency also depends on the inclination angle $i$
between the galactic disk and the ICM wind direction (Vollmer et al. 2001).
Since the H{\sc i} observations indicate that
stripping occurs more face-on, we do not consider edge-on stripping.
We made 3 simulations with 3 different 
inclination angles between the galaxy's disk and the ICM wind direction:
(i) $i=45^{\circ}$, (ii) $i=60^{\circ}$, and (iii) $i=75^{\circ}$.
An inclination of $i=0^{\circ}$ means that the galactic disk is
parallel to the ICM wind direction.

\begin{figure}
	\resizebox{\hsize}{!}{\includegraphics{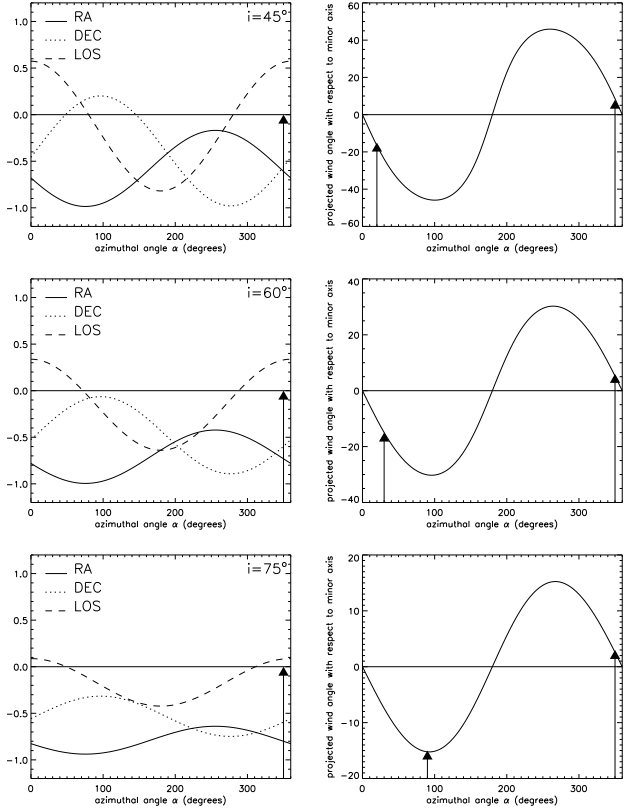}}
	\caption{Inclination angle between the galactic disk and the ICM wind $i=45^{\circ}$,
	  $i=60^{\circ}$, and $i=75^{\circ}$.
	  Left panels: components of the 3D ICM wind direction as 
	  a function of the azimuthal viewing angle $\alpha$.
	  Right panels: projected wind direction with respect to the galaxy's minor
	  axis (counted counter-clockwise) as a function of $\alpha$.	 
	  The arrows indicate the best fit model $\alpha=350^{\circ}$ and
	  the azimuthal viewing angle for a projected wind angle of $-15^{\circ}$.
	} \label{fig:angles}
\end{figure}

The last free parameter, the azimuthal viewing angle, is chosen
in a way to fit the observed H{\sc i} distribution and to reproduce
the positive line-of-sight component of the wind direction (the galaxy is moving away 
from the observer). In Figs.~\ref{fig:angles} we show the components of the 3D ICM wind direction 
(left panels)
and the projected wind direction as a function of the azimuthal viewing angle $\alpha$ (right panels).
The maximum of the line-of sight component of the ICM wind is 
largest for $i=45^{\circ}$ and smallest for $i=75^{\circ}$.
The wind components along the right ascension and declination are
all negative, i.e. the ICM wind blows from the south east.
Whereas the two components are roughly equal for $i=45^{\circ}$ and $i=60^{\circ}$,
the RA components is smaller than the DEC components for $i=75^{\circ}$.
As the  maximum of the line-of sight component the maximum of the projected wind direction 
also varies with $i$. It is maximum for $i=45^{\circ}$ and minimum for $i=75^{\circ}$.
In all cases, an azimuthal viewing angle of $\alpha=350^{\circ}$ leads to 
a positive projected wind direction smaller than $6^{\circ}$.
However, if one assumes a given projected wind direction of $-15^{\circ}$,
the corresponding azimuthal viewing angles vary: $\alpha(45^{\circ})=20^{\circ}$,
$\alpha(60^{\circ})=30^{\circ}$, $\alpha(75^{\circ})=90^{\circ}$.

\begin{figure*}
	\resizebox{\hsize}{!}{\includegraphics{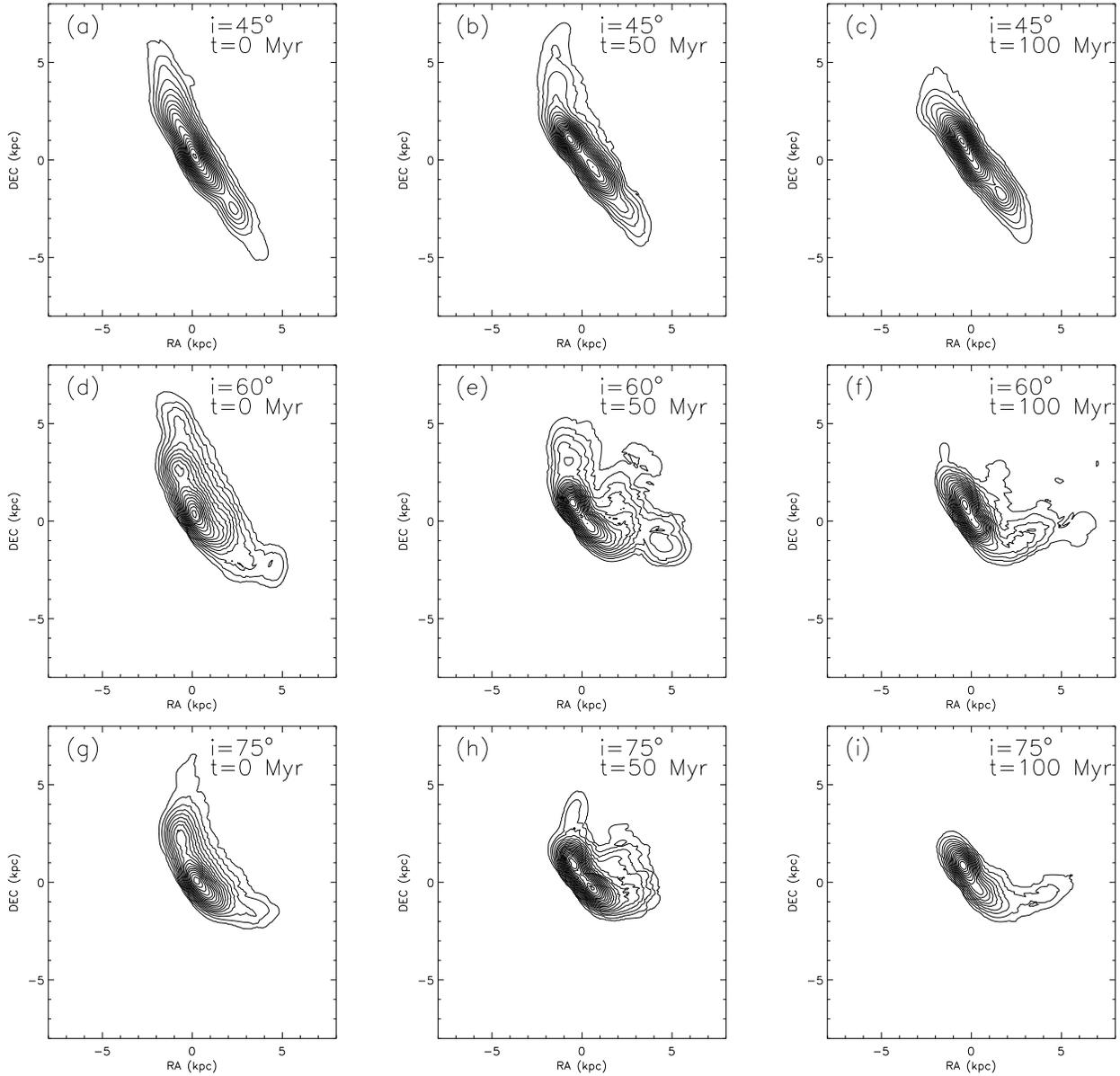}}
	\caption{Model snapshots for different simulations.
	 The ram pressure profile for all simulations is
	 given by Eq.~\ref{eq:rps}. The inclination angle $i$
	 between the orbital and the galaxy's disk plane and
	 the timestep of the snapshot are varied.
	 Left column: $t=0$~Myr. Middle column: $t=50$~Myr.
	 Right column: $t=100$~Myr. Upper row: $i=45^{\circ}$.
	 Middle row: $i=60^{\circ}$. Lower row: $i=75^{\circ}$.
	} \label{fig:mom0s}
\end{figure*} 

None of the simulations shows significant extraplanar gas before $t=0$~Myr.
In Fig.~\ref{fig:mom0s} we show 3 timesteps 0~Myr, 50~Myr, and 100~Myr
after the ram pressure maximum.
The simulation with an inclination angle between the galaxy's disk and
the ICM wind direction of $i=45^{\circ}$ does not lead to the observed stripping
radius neither shows significant extraplanar gas.
The simulation with $i=75^{\circ}$ (Fig.~\ref{fig:mom0s}a-c) leads to the correct stripping
radius and shows extraplanar gas at $t=50$~Myr (Fig.~\ref{fig:mom0s}h)
and $t=100$~Myr (Fig.~\ref{fig:mom0s}i).
However, the column density of this extraplanar gas is smaller
compared to the H{\sc i} observations. In particular, the observed detached south
western extraplanar H{\sc i} maximum is missing.
Only the simulation with $i=60^{\circ}$ at $t=50$~Myr (Fig.~\ref{fig:mom0s}e)
can reproduce the observed
H{\sc i} distribution. About $50$~Myr later the column density of 
the northern extraplanar gas has already decreased significantly and the
shape of the southern extraplanar gas has changed (Fig.~\ref{fig:mom0s}f).

\begin{figure}
	\resizebox{\hsize}{!}{\includegraphics{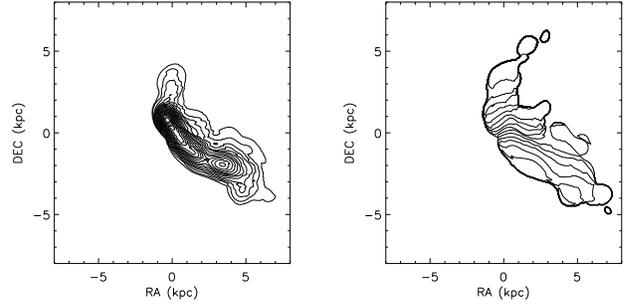}}
	\caption{Snapshot at $t=40$~Myr of the simulation with $i=60^{\circ}$.
	 The azimuthal viewing angle is $\alpha=30^{\circ}$.
	 Left panel: gas distribution. Right panel: velocity field.
	} \label{fig:newangle}
\end{figure}

As a conclusion, we rule out the $i=75^{\circ}$ simulations, because the
galaxy's small line-of-sight velocity component (Fig.~\ref{fig:angles})
would lead to galaxy and/or ICM velocities which are beyond any reasonable values
($>4000$~km\,s$^{-1}$). We also rule out the $i=45^{\circ}$ simulations, because
it can not reproduce the observed extraplanar gas. We are thus left with the 
$i=60^{\circ}$ simulation. Recent HST observations of NGC~4522 (Kenney et al., in prep.)
yield a projected wind direction of $\sim -15^{\circ}$. The corresponding
azimuthal viewing angle is $\alpha=30^{\circ}$. Using this $\alpha$ we realized that the
observations can be best reproduced with $t=40$~Myr instead of $t=50$~Myr used before
(Fig.~\ref{fig:newangle}). Whereas the H{\sc i} gas distribution is reproduced in
a satisfactory way, the velocity isocontours of the western extraplanar gas
are different from the observed ones (Fig.~\ref{fig:moments}).

\begin{figure}
	\resizebox{\hsize}{!}{\includegraphics{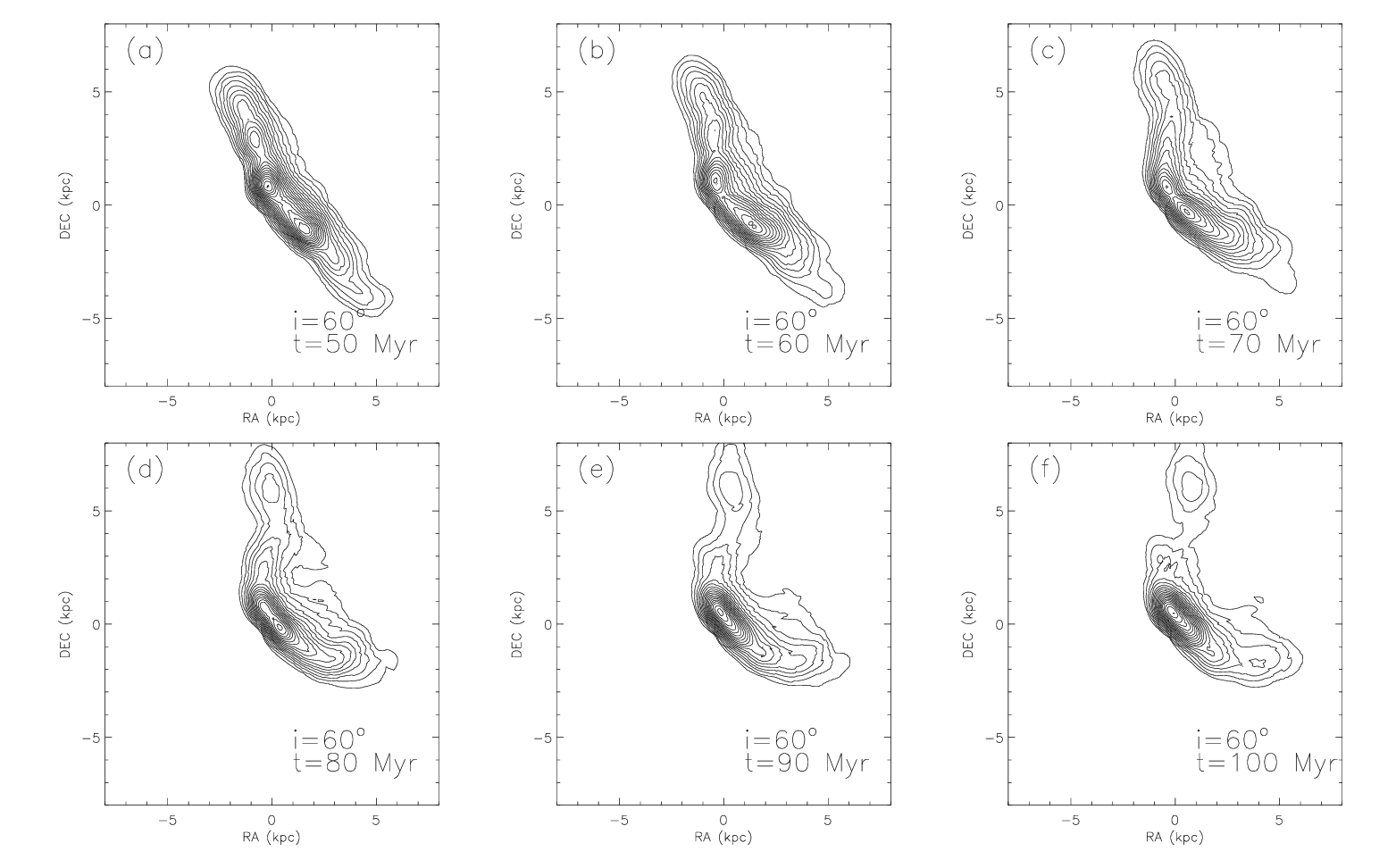}}
	\caption{Snapshots of the simulation using a step function for the
	  temporal ram pressure profile with $i=60^{\circ}$.
	 The azimuthal viewing angle is $\alpha=350^{\circ}$.
	} \label{fig:stepfunction}
\end{figure}

As an additional test we approximated the temporal ram pressure profile 
with a step function, i.e. the galaxy is moving abruptly into a region
of high intracluster medium density and/or velocity. The step occurs at $t=0$~Myr and
the subsequent ram pressure is assumed to be constant at a level of the maximum
ram pressure of the Lorentzian (Eq.~\ref{eq:rps}).
Fig.~\ref{fig:stepfunction} shows the ISM evolution for $i=60^{\circ}$.
None of the timesteps show the observed distribution of the
extraplanar high column density gas (Fig.~\ref{fig:moments}).
We therefore exclude a constant ongoing ram pressure.

Therefore, our best fitting model has the following parameters (see Eq.~\ref{eq:rps}):
$p_{\rm max}=2000$~cm$^{-3}$(km\,s$^{-1}$)$^{2}$, $t_{\rm HW}$=80~Myr,
$i=60^{\circ}$, and $t=50$~Myr. The associated uncertainties are $\Delta i \sim 10^{\circ}$,
$\Delta \alpha \sim 20^{\circ}$, and $\Delta t \sim 10$~Myr.
Subsequently, the radial velocity
of NGC~4522 represents $33$\,\% of its total velocity within a static ICM.
The total velocity of NGC~4522 with respect to a static ICM is thus
$v_{\rm N4522} \sim 3500$~km\,s$^{-1}$. Together with a local ICM density of
$n_{\rm ICM}=10^{-4}$~km\,s$^{-1}$ this is consistent with our assumed peak
ram pressure of 2000~cm$^{-3}$(km\,s$^{-1}$)$^{2}$ within a factor of 1.6.

\begin{figure}
	\resizebox{5cm}{!}{\includegraphics{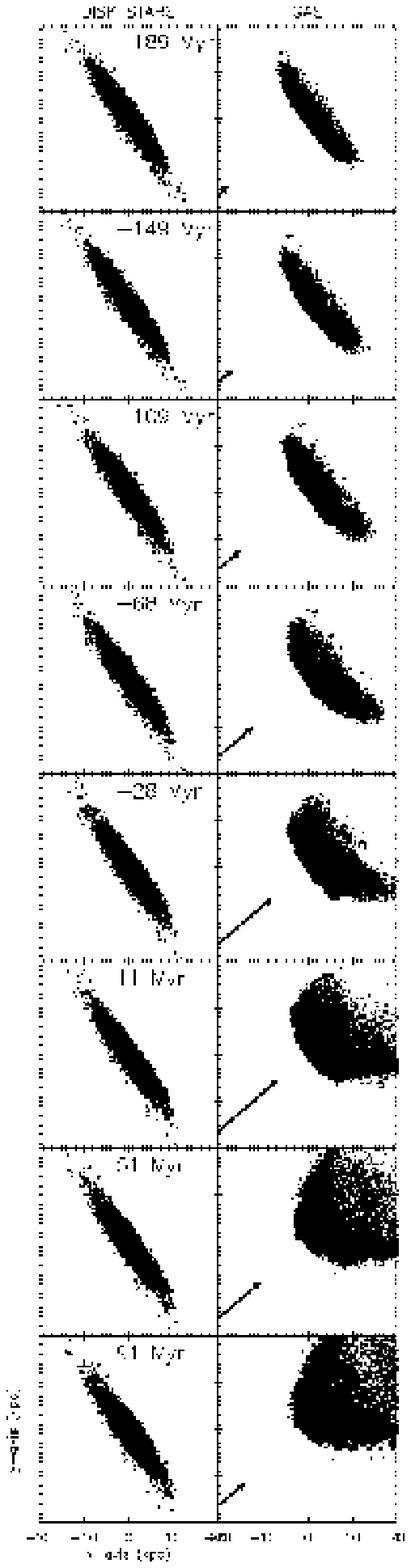}}
	\caption{Evolution of the model stellar (1st column)
	and gas disk (2nd column). 
	The major axis position angle and inclination of NGC~4522 
	are $PA=214^{\circ}$ and $i=80^{\circ}$, respectively. The arrow
	indicates the direction of ram pressure, i.e. it is opposite to
	the galaxy's velocity vector, and its size is proportional
	to $\rho v_{\rm gal}^{2}$. Maximum ram pressure occurs at
	$t=0$~Myr. The timestep of each snapshot is marked in each panel
	showing the stellar disk.
	} \label{fig:evolution}
\end{figure}

The large scale evolution of this NGC~4522 simulation can be seen in Fig.~\ref{fig:evolution}.
The ICM wind begins to drive out the ISM of NGC~4522 only at $\sim -100$~Myr. 
The gas is expelled to the west. At the
first stage of stripping ($t \sim -30$~Myr) the extraplanar gas of highest surface
density is found in the south-west. At $t \sim 10$~Myr both, the south-western and
north-western parts of the extraplanar gas, show a relatively high column density.
After the ram pressure maximum has occurred ($t=0$~Myr) the surface density of the
extraplanar gas west to the galaxy center has significantly increased.

\begin{figure*}
	\resizebox{\hsize}{!}{\includegraphics{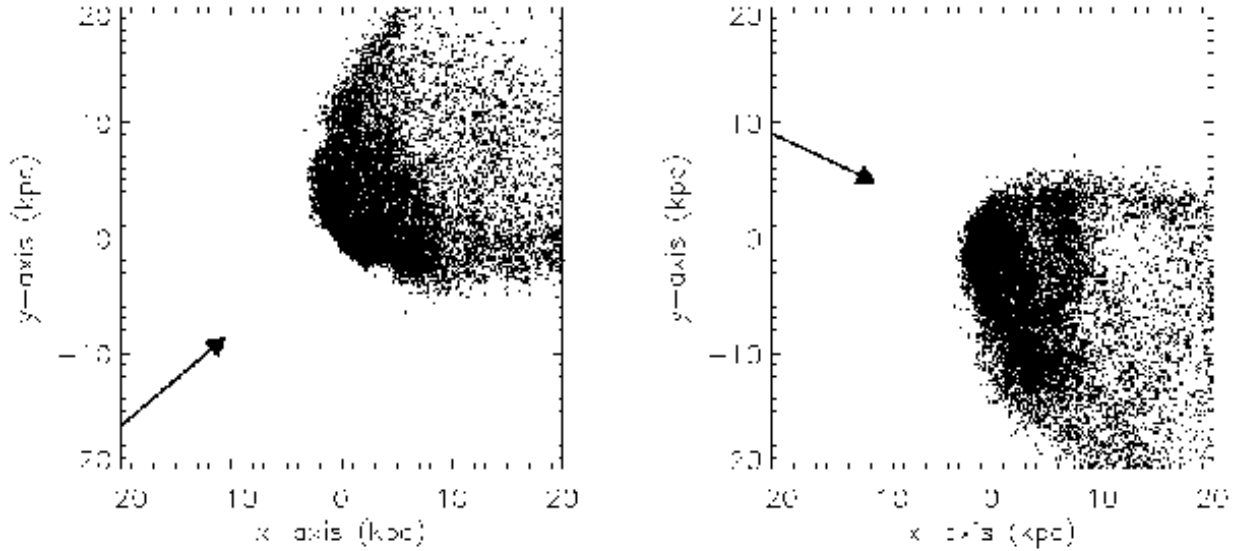}}
	\caption{Comparison snapshot of the gas distribution of NGC~4522. Left panel:
	  view of the disk using the observed position angle and 
	  inclination of NGC~4522. Right panel: the disk seen face-on.
	  The arrow indicates the direction of the wind,
	  i.e. opposite to the galaxy's motion.
	} \label{fig:snapshot}
\end{figure*}

According to these simulations we observe the galaxy $\sim 50$~Myr after the ram 
pressure maximum.
In the two different scenarios described above this means that (i) the
galaxy encountered the intracluster medium of highest density $\sim 50$~Myr ago or
(ii) the galaxy just leaves the region of maximum ICM velocity.
The final gas distribution is shown in Fig.~\ref{fig:snapshot}.
The extraplanar gas has a complex asymmetric three-dimensional structure.
A part of the ISM initially located at galactic radii between 3 and 5~kpc is stripped in
a ring-like configuration, located up to $\sim 4$~kpc above the stellar disk.

\section{Comparison with VLA H{\sc i} observations \label{sec:comparison}}

In this section the model snapshot of Fig.~\ref{fig:snapshot} is compared with
the VLA H{\sc i} observations of Kenney et al. (2004). For this purpose we
assume that only gas with a volume density greater than $\sim 10$~cm$^{-3}$
is in form of neutral hydrogen. This corresponds to the high end
of the densities of the warm ionized medium (Boulares \& Cox 1990).
The lower density gas is assumed to be 
ionized. The comparison between the model and observations is done based on
moment maps (Sect.~\ref{subsec:moments}) and a position-velocity diagram
(Sect.~\ref{subsec:pv}).

\subsection{The moment maps \label{subsec:moments}}

\begin{figure}
	\resizebox{\hsize}{!}{\includegraphics{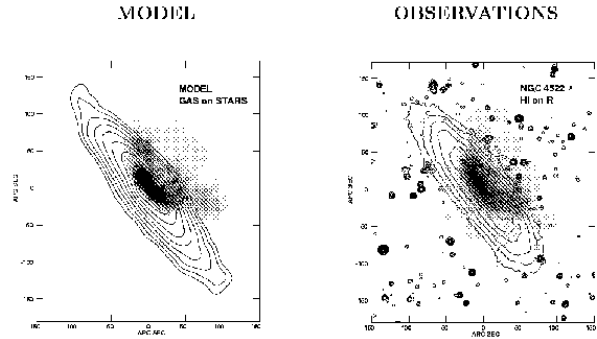}}
	\caption{Stellar (contours) and gas (greyscale) distribution of NGC~4522.
	Left panel: model. Right panel: H{\sc i} observations (Kenney et al. 2004).
	} \label{fig:mom0stars}
\end{figure}

Since ram pressure selectively affects the gas, the model and observed stellar disks
are symmetric whereas the model and observed gas distributions are highly
asymmetric (Fig.~\ref{fig:mom0stars}), i.e. the ISM is pushed to the west
of the galactic disk.

\begin{figure*}
	\resizebox{16cm}{!}{\includegraphics{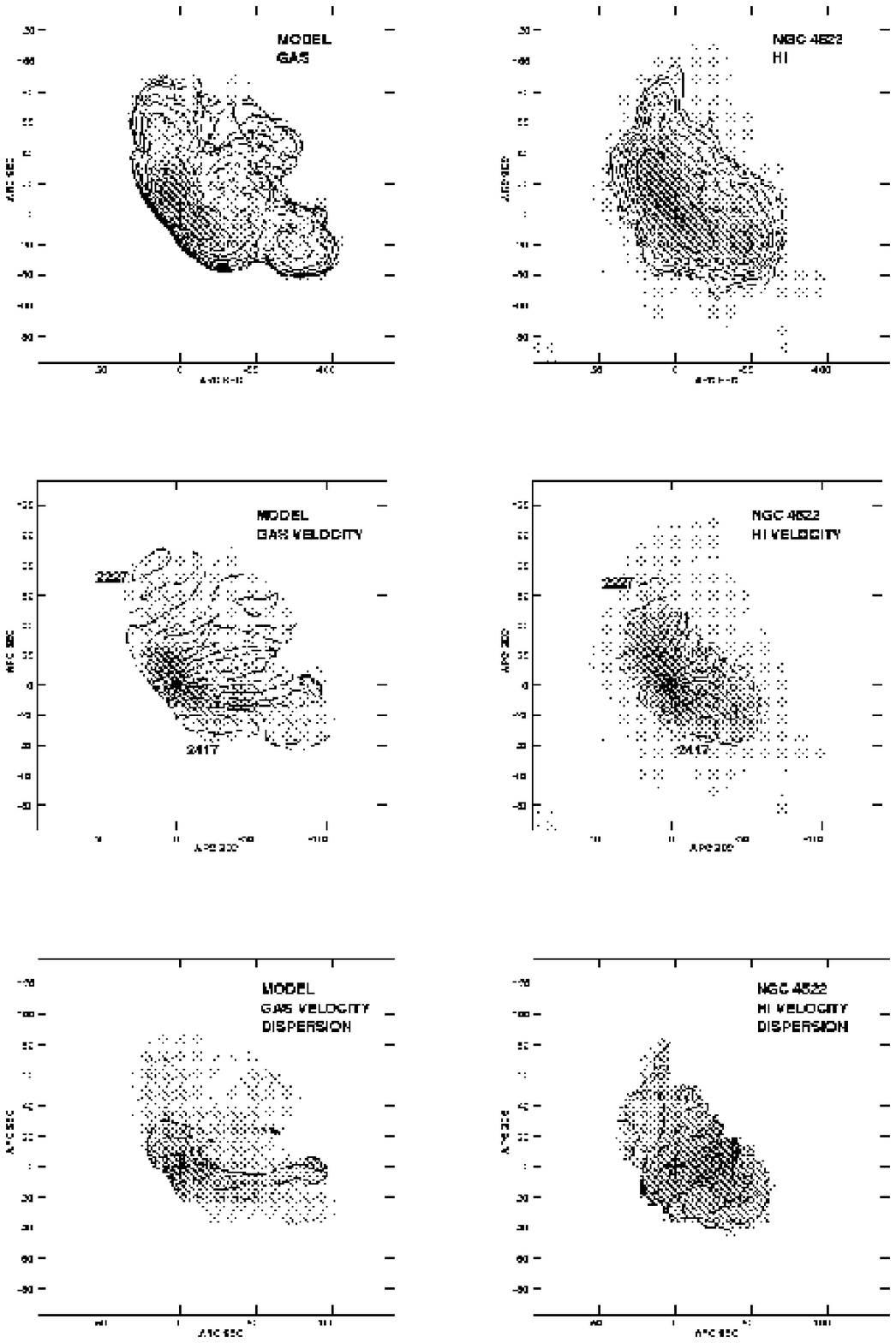}}
	\caption{Moment maps of NGC~4522. Left panels: model. Right panels:
	H{\sc i} observations (Kenney et al. 2004).
	The same contouring scheme is used for both model and data.
	Upper panels: gas distribution; contours spaced by $\sqrt{2}$. 
	Middle panel: velocity field; contours spaced 10~km\,s$^{-1}$.
	Lower panel: velocity dispersion; contours: 15, 20, 25, 30, 35, 40, 45~km\,s$^{-1}$.
	} \label{fig:moments}
\end{figure*}

The H{\sc i} distribution (moment 0), velocity field (moment 1), and velocity
dispersion (moment 2) are shown in Fig.~\ref{fig:moments}.
The following observed properties of the H{\sc i} data are reproduced:
\begin{itemize}
\item
the model H{\sc i} deficiency is 0.5 compared to the observed value of $0.6 \pm 0.2$,
\item
the gas disk is truncated at a radius of about 3~kpc,
\item
within the disk more gas is found in the northern part than in the southern part,
\item
the existence of a high surface density extraplanar gas component to the west
of the galactic disk,
\item
the extraplanar gas distribution shows two maxima; the south-western
maximum is more prominent than the north-western one,
\item
the velocity field of the extraplanar gas is relatively regular,
\item
the extraplanar gas has less extreme velocities than the nearby disk emission,
i.e. the velocity contours curve away curve away from the minor axis.
\end{itemize}
Thus, the overall agreement between the model and observed first two moments
(gas distribution and velocity field) is satisfactory.
Whereas 40\% of the observed H{\sc i} mass resides in the extraplanar 
component, the model yields a fraction of 65\%.
This difference is likely due to the unknown difference between the initial gas distribution
of the model and/or due to the fact that we detect only a fraction of the stripped gas
due to ISM expansion, heating and ionization.

The model and observed velocity dispersions are different.
In the observations, the peak H{\sc i} dispersion is not at the nucleus,
but in the extraplanar gas, $\sim 30''$ west of the nucleus.
Even within the disk, the maximum in the observed dispersion is offset
from the nucleus by $\sim 10''$.
Our model shows a maximum in the center, a 
region of marginally larger linewidth to the north, and
a region of larger linewidth within the south western
extraplanar gas. The central model peak is due to the model rotation
curve which is somewhat steeper than the observed rotation curve.
Thus, although we find a small enhancement of the linewidth
in the south western extraplanar gas, its linewidth is a factor of 3 
smaller than the observed one.

\subsection{Position-velocity diagram \label{subsec:pv}}

\begin{figure*}
	\resizebox{\hsize}{!}{\includegraphics{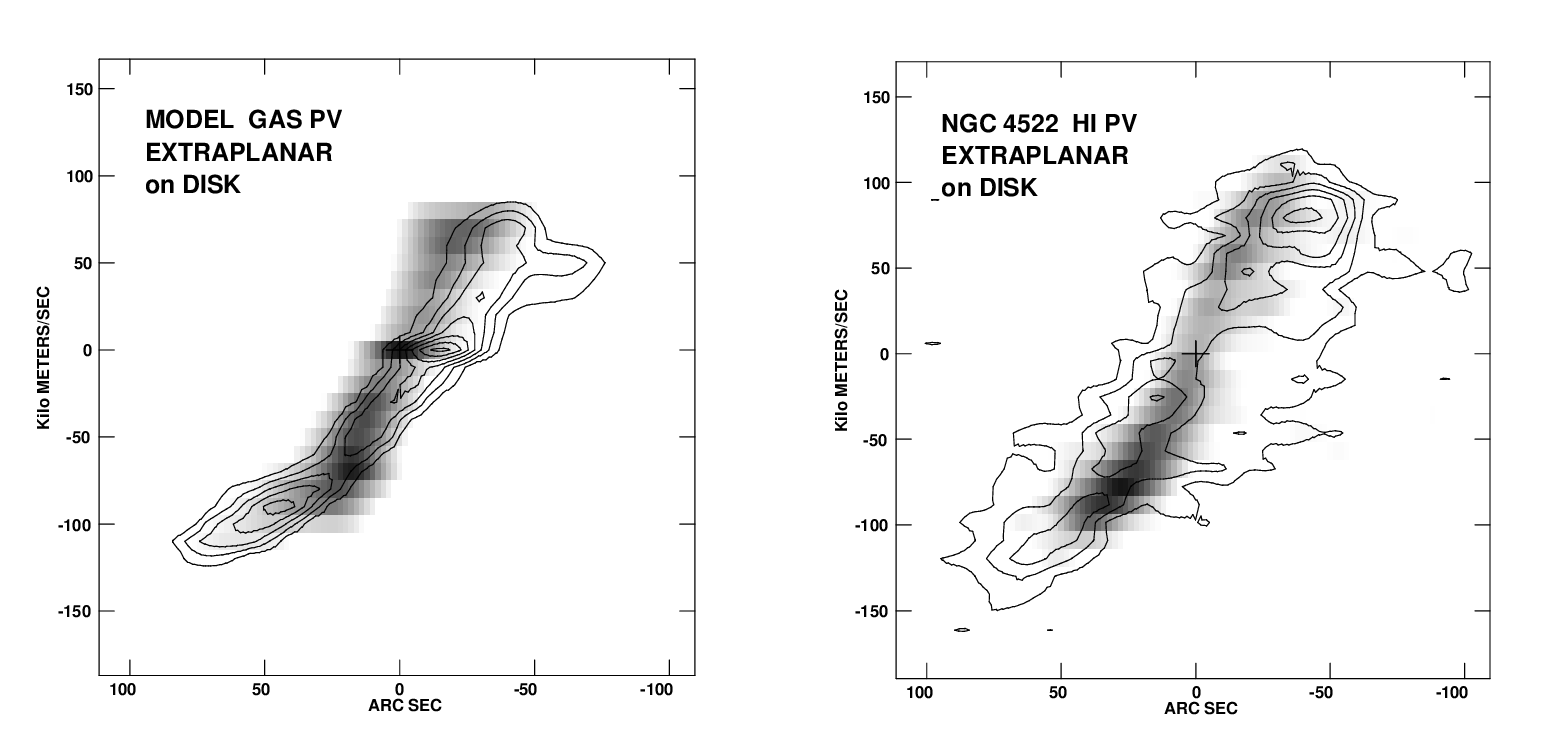}}
	\caption{Major axis position-velocity diagram for the
	  disk (greyscale) and extraplanar (contours) component.
	  Left panel: model. Right panel: H{\sc i} observations
	  (Kenney et al. 2004). Contour levels are 1, 2, 3, 4, 5, 6 
	  times the peak value.
	} \label{fig:pvdiagram}
\end{figure*}

The position-velocity diagram along the major axis for the
disk and extraplanar regions are shown in Fig.~\ref{fig:pvdiagram}.
The disk region is defined as the region between $-20''$ and $+11''$
from the major axis. The extraplanar region is defined as
the region between $+11''$ and $+50''$ from the major axis.
The following observed characteristics are reproduced by the model:
\begin{itemize}
\item
the disk gas kinematics are relatively symmetric and normal,
despite the strong north-south asymmetry of the gas distribution,
\item
the extraplanar gas has lower (blueshifted) velocities with respect
to the disk gas.
\end{itemize}
On the other hand, our model does not reproduce the observed large linewidth
(up to $150$~km\,s$^{-1}$ FWZI).

\section{MHD simulations \label{sec:mhd}}

\subsection{The model \label{subsec:mhdmodel}}

Otmianowska-Mazur \& Vollmer (2003) studied the evolution of the
large scale magnetic field during a ram pressure stripping event.
They calculated the magnetic field structure by solving the induction equation 
on the velocity fields produced by the dynamical model.
The polarized radio continuum emission has been calculated by assuming
a Gaussian spatial distribution of relativistic electrons. This procedure
allowed them to study the evolution of the observable polarized radio
continuum emission during a ram pressure stripping event.

We apply the same procedure as Otmianowska-Mazur \& Vollmer (2003)
on a similar ram pressure stripping event (Sect.~\ref{sec:model}).
The Zeus3D code  (Stone \& Norman 1992a and b) is used to solve
the induction equation: 
\begin{equation}
{\partial\vec{B}/\partial t=\hbox{rot}(\vec{v}\times\vec{B})
 -\hbox{rot}(\eta~\hbox{rot}\vec{B})}
\label{eq:inductioneq}
\end{equation}
where $\vec{B}$ is the magnetic induction, $\vec{v}$ is the large-scale
velocity of the gas, and $\eta$ is the coefficient of a turbulent diffusion.
We use a physical diffusion of $\eta = 5 \times 10^{25}$~cm$^{2}$s$^{-1}$
(Elstner et al. 2000).
The estimated numerical diffusion is an order of magnitude smaller and
thus does not affect our simulations. We do not use an $\alpha$ dynamo. 
The initial magnetic field is purely toroidal with a strength of 10~$\mu$G.

The induction equation is solved on a rectangular coordinates ($XYZ$).
The number of grid points used is 171x171x71 along the $X$, $Y$ and $Z$ axis, 
respectively. This corresponds to the 
grid spacing of 200~pc in the galactic plane and
of 300~pc in the $Z$ direction, resulting in a size of the modeled box of
34.2~kpc~$\times$~34.2~kpc~$\times$~21.3~kpc. Since the N-body code is discrete
whereas the MHD code is using a grid, we have to interpolate the
discrete velocities on the grid. This is done using a method
known as ``Kriging'' with a density-dependent smoothing length
(Isaaks \& Srivastava 1989). It turned out that we
had to use a large smoothing length  to suppress the noise 
in the velocity field of the outer disk, which is due to a small, local particle density.
In this way we avoid numerical artifacts of the magnetic field distribution
at the outer disk.
As the spline interpolation used in Otmianowska-Mazur \& Vollmer (2003),
this has the consequence that the velocity field at the edge of the gas
distribution is more extended than the gas distribution itself.
Since there are gradients in this velocity field, induction leads to a 
magnetic field which extends beyond the edge of the gas distribution.
This affects the polarized emission beyond the gas distribution, but
not inside, which is what we are interested in.

\begin{figure*} 
	\resizebox{12cm}{!}{\includegraphics{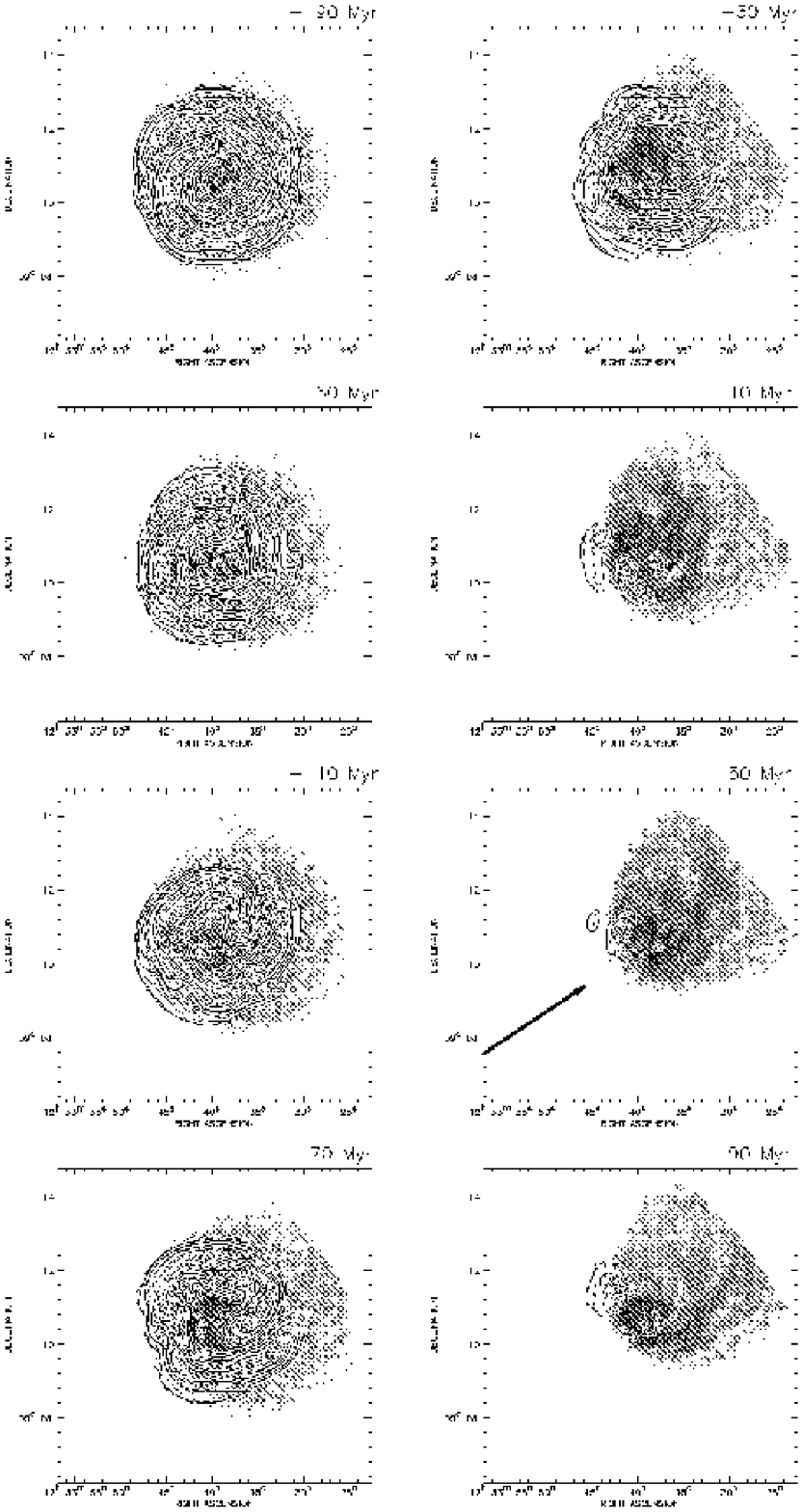}}
	\caption{Evolution of the polarized radio continuum emission. Time steps
are the same as in Fig.~\ref{fig:evolution}. The disk is seen face-on and
rotates clockwise. Contours: polarized radio continuum emission. The magnetic field
vectors are superimposed onto the gas surface density. 
The timestep is indicated on top of each snapshot. The assumed beamsize for the
polarized radio continuum emission is $20''$. The arrow in the last panel
indicates the direction of the ram pressure wind.
	} \label{fig:pievolution}
\end{figure*}

The evolution of the polarized radio continuum emission without Faraday rotation
is presented in Fig.~\ref{fig:pievolution}. The timesteps are the same as in
Fig.~\ref{fig:evolution}. The disk rotates clock-wise.
The gas surface density, which is smoothed
to a resolution of $\sim 100$~pc, is shown in greyscales,
the polarized radio continuum emission as contours, and the magnetic field vectors
projected on the plane of the sky as lines.
We assume a Gaussian distribution of relativistic electrons in $R$ and $z$ directions:
$n_{\rm rel}=n_{0}\,\exp{\big(-(r/r_{\rm R/z})^2\big)}$, where $r_{\rm R}=5$~kpc and $r_{\rm z}=0.5$~kpc.
This translates into a FWHM of 8.3~kpc and 1.2~kpc, respectively.
By assuming this smooth distribution we imply no equipartition between the energy
densities of total cosmic rays and total magnetic field. 
The galaxy moves to the lower left corner, i.e. the ram pressure wind comes
from the this direction. The ram pressure maximum occurs at $t=0$~Myr.
As already seen in Sect.~\ref{sec:model} ram pressure begins to push the gas to the
upper right corner at $t \sim -100$~Myr. The straight cut-off of the gas distribution 
in the upper right quadrant is due to the edge of our computational volume.

At $t > -100$~Myr the gas is compressed at the lower left
side of the galaxy. At the same time most of the gas initially located in the outer
disk ($R > 3$~kpc) is pushed out of the disk plane (Fig.~\ref{fig:evolution}).
At $t > 0$~Myr a prominent maximum of the polarized radio continuum emission
forms in the lower left quadrant. Due to our large
density-dependent smoothing length, a part of the large-scale
magnetic field in the lower left quadrant of the galaxy does not follow the compression
and stays at large galactic radii in the disk plane (see also Otmianowska-Mazur \& Vollmer 2003). 
Therefore, the distribution of polarized intensity extends
beyond that of the gas which is mostly due to our smoothing algorithm.
However, a decoupling of the large-scale magnetic field from the gas flow is
not entirely excluded (see .e.g. Soida et al. 2001). 
Recently, Beck et al. (2005) could show that the regular magnetic field can decouple
from the cold gas.
The maximum of polarized emission within the gas disk in the lower left direction is real
and due to gas compression.
As observed in Otmianowska-Mazur \& Vollmer (2003) after maximum compression at $t=0$~Myr this inner 
maximum is taken along with rotation towards the upper left part of the image ($t > 20$~Myr).
At $t=50$~Myr it is located to the left of the galaxy center, at $t=90$~Myr it is located
on the upper left part of the gas disk.
Since the FWHM of the relativistic electron distribution in the 
vertical direction is 1.2~kpc, the magnetic field taken away vertically with the gas is no longer 
observable in polarized radio continuum emission because of the low density
of relativistic electrons there. 

The evolution of the polarized radio continuum emission (Fig.~\ref{fig:pievolution})
is qualitatively similar to that of Otmianowska-Mazur \& Vollmer (2003), where an inclination
angle between the disk and the ICM wind direction of $i=20^{\circ}$ was assumed.
Thus, the Otmianowska-Mazur \& Vollmer (2003) simulation corresponded to a more edge-on
stripping, whereas in our simulation ram pressure stripping occurs more face-on 
($i=60^{\circ}$). Consequently, the gas compression is less pronounced in the
present case. However, as in the Otmianowska-Mazur \& Vollmer (2003) simulations
we observe that shortly after the timestep of maximum ram pressure the maximum of 
polarized radio continuum emission within the compressed gas disk is moving to the 
upper part of the image due to galactic rotation. However, ram pressure is still
quite high and the ISM is still leaving the galactic disk at $t=50$~Myr.
This is preferentially happening on the side where galactic rotation and
the ICM wind are parallel, i.e. in the north east.

As a conclusion, the polarized radio continuum emission outside the gas disk
at timesteps $t > -20$~Myr is most probably due to our 
large density-dependent smoothing length 
due to the intrinsic difficulties of interpolating and extrapolating the 3D 
velocity field over a sharp edge of the gas distribution where no
particles are found beyond a certain radius. In this case an interpolation and extrapolation 
with a density dependent smoothing length can always give rise to
artifical velocity gradients which lead to an enhancement of the magnetic field.
The here employed ``Kriging'' method with a density dependend smoothing length
turned out to be a good compromise between the need of a smooth velocity and a
minimum of artificial velocity gradients at the edge of the gas distribution.

However, the maximum
of polarized radio continuum emission inside the compressed gas disk
is real. Since we know the cause of polarized emission extending beyond the gas disk
and since this feature and the real inner maximum of polarized emission are  
very close in space and since 
we did not want to modify the model by hand with a risk of introducing uncontrollable 
artifacts, we did not attempt to remove the part of the large-scale magnetic field
extending beyond the gas disk.
We interpret the observed maximum of polarized radio continuum emission within the gas
disk as being due to gas compression.

\subsection{Comparison with VLA polarized radio continuum emission \label{subsec:PI}}

The distribution of polarized radio continuum emission is now projected using the
position and inclination angles of NGC~4522, $t=50$~Myr, and the azimuthal angle as 
in Sect.~\ref{sec:model}.
\begin{figure*}
	\resizebox{\hsize}{!}{\includegraphics{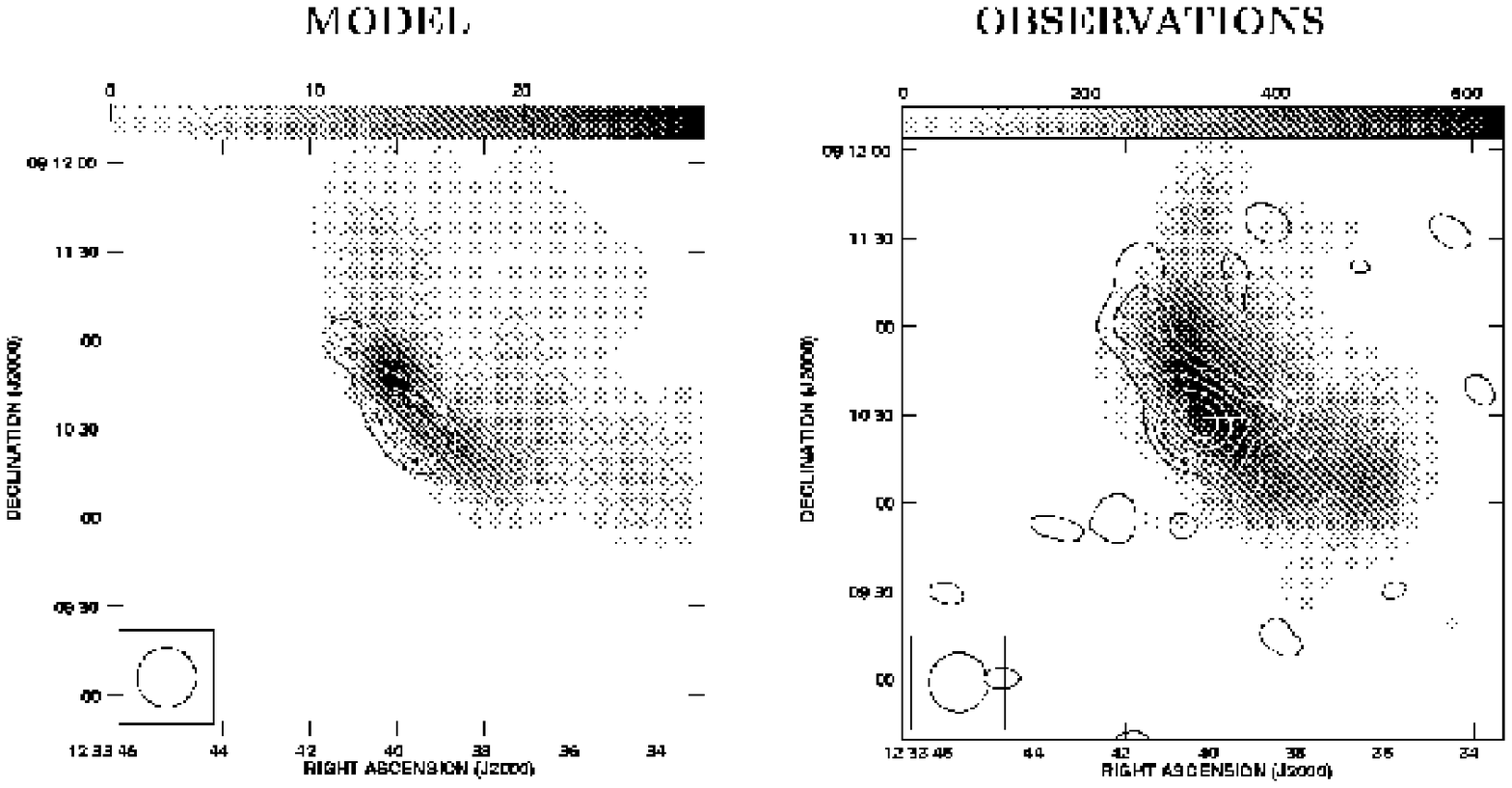}}
	\caption{Greyscale: H{\sc i} gas distribution
	with a resolution of $\sim$20$''$. Contours: polarized
	radio continuum emission. Left panel: model. The FWHM of the relativistic 
	electron distribution in the radial and vertical directions
	are 8.3~kpc and 1.2~kpc, respectively. Right panel:
	observations (Vollmer et al. 2004).
	} \label{fig:gaspi} 
\end{figure*} 

The resulting model polarized radio continuum emission distribution
and the model magnetic field vectors on the model gas distribution are shown in 
Fig.~\ref{fig:gaspi} together with our 6~cm VLA observations (Vollmer et al. 2004).
As our observations, the model shows a polarized radio continuum maximum to
the east of the galaxy center and a weaker extension to the north.
The observed offset of the polarized emission maximum 
to the east translates the fact that the inclination angle of the
galaxy is such that the eastern side of the disk is the compressed side
which is the far side of the galaxy.
Since the polarized emission extending beyond the edge of the
gas distribution and the real compressed large-scale
magnetic field are contiguous, the shown distribution of polarized radio continuum emission
reflects that of the compressed component. The observed elongation of the polarized radio
continuum emission can be explained as follows: due to gas compression a polarized emission
maximum is created in the south-east. Shortly after the occurrence of the ram pressure maximum,
the large-scale magnetic field follows rotation to the north-east.
Thus, the observed polarized radio continuum emission corroborates our previous
finding (based on the comparison between the model and the H{\sc i} data) that
we observe NGC~4522 after the occurrence of maximum ram pressure (Sect.~\ref{sec:bestfit}).

\section{Discussion \label{sec:discussion}}

\subsection{Stripping of the cold and warm H{\sc i}}

For face-on stripping, our simulations are consistent with the
Gunn \& Gott (1972) ram pressure estimate.
Our dynamical model reproduces qualitatively and quantitatively the observed H{\sc i} gas distribution,
velocity field (Fig.~\ref{fig:moments}) and polarized radio continuum emission distribution
(Fig.~\ref{fig:gaspi}). However, it fails in reproducing the observed large H{\sc i} linewidths
in the extraplanar component (Fig.~\ref{fig:pvdiagram}). 
This is most probably due to a lack of gas physics in the model.
The large observed linewidth of $\sim 100$~km\,s$^{-1}$ cannot be thermal, because
this would translate into temperatures where the gas is ionized.
Thus, they are due to directed, chaotic or turbulent gas motions.
Since the H{\sc i} line profiles tend to smaller velocities, i.e. to velocities closer to that of
the Virgo cluster, one possible explanation is that a part of the atomic gas is stripped
more efficiently than the rest. Since the column density of the low velocity gas
is small, we speculate that it is a low density warm ($T \sim 8000$~K) atomic gas phase
which is stripped more efficiently.
In this scenario the large linewidth is due to directed motions.
In a second scenario ram pressure leads to shocks in the interstellar medium which
increase their chaotic or turbulent velocity dispersion dramatically.
Again, as the column density of the low velocity gas is small, this might only concern 
the low density warm component of the atomic gas which has a larger
volume filling factor than the cold atomic gas ($T \sim 100$~K). 
Both scenarios can explain the observed large H{\sc i} linewidths in the extraplanar component.
In both scenarios it is the warm diffuse atomic gas phase which produces the
large linewidths.

\subsection{The stripping efficiency}

As already mentioned in Sect.~\ref{sec:model} our knowledge about the orbit of NGC~4522
within the Virgo intracluster medium is poor. Its high radial velocity with respect
to the cluster mean and its large distance from the cluster center exclude a
simple radial orbit. Kenney et al. (2004) estimated ram pressure due to a static 
intracluster medium at the projected location of NGC~4522. Applying the Gunn \&
Gott (1972) criterion and assuming an ICM
density of $n_{\rm ICM}=10^{-4}$~cm$^{-3}$ and a galaxy velocity of $1500$~km\,s$^{-1}$,
they found that the gravitational restoring force for a gas cloud with a surface density
of $10$~M$_{\odot}$pc$^{-2}$ is an order of magnitude higher
than the force due to ram pressure. This led them to the conclusion that ram pressure is higher than
the ``standard value''. This could in principle be due to a higher galaxy velocity or
a higher stripping efficiency $\xi$ with $p_{\rm ram}=\xi \rho (\vec{v_{\rm gal}}-\vec{v_{\rm ICM}})^2$ 
(see e.g. Roediger \& Hensler 2005).

Our simulation yields a projected ICM wind direction to the north east which 
is roughly aligned with the galaxy's minor axis ($6^{\circ}\pm 20^{\circ}$; 
Fig.~\ref{fig:evolution}). This is 
corroborated by the recent results of Kenney et al. (in prep.) analysing dust extinction features in
deep HST images of NGC~4522. They found gas clouds beyond the stripping radius
in the direction of the ICM wind. Beyond the main gas/dust truncation radius,
they found several elongated dust clouds, all with similar position angles of
$-17^{\circ} \pm 3^{\circ}$ with respect to the minor axis, likely indicating
the projected ICM wind direction.
Moreover, our simulations yield an inclination angle between the disk and the 
ICM wind direction of $i=60^{\circ} \pm 10^{\circ}$ (more face-on stripping). 
Thus the wind angle is relatively face-on despite the fact that the galaxy us nearly
edge-on and has a high line-of-sight velocity, implying that there must be a large 
component of the ICM wind in the plane of the sky. Based on our model the total 
velocity of the galaxy relative to a static ICM is $\sim 3500$~km\,s$^{-1}$ (see Sect.~\ref{sec:bestfit}).
With this velocity the Gunn \& Gott criterion is approximately fulfilled and the stripping 
efficiency is close to one.

\subsection{Static or moving ICM}

The comparison between H{\sc i} observations and our model
(Sect.~\ref{sec:comparison}) suggests that ram pressure acting on NGC~4522 is close to 
its maximum but has already passed it. 
The comparison between the observed and modeled polarized radio continuum emission
confirms that the maximum has occurred recently ($\sim 50$~Myr ago), because the polarized
radio continuum emission extends to the north-east, in the direction of galactic rotation.

Thus, if the galaxy is crossing the cluster with a very high velocity ($\sim 4000$~km\,$^{-1}$)
and the ICM is static and smooth,
NGC~4522 has just passed the point of highest intracluster medium density, i.e. it 
has just passed the point of smallest distance to M87.
On the other hand, if the intracluster medium is moving due to the infall
of the M49 group of galaxies as suggested by Kenney et al. (2004) and Vollmer et al. (2004),
the galaxy has just passed the region of highest intracluster medium velocities.
In this case ram pressure is given by Eq.~\ref{eq:icmwind}.
If the main component of $\vec{v_{\rm gal}}$ is radial and thus 
$v_{\rm gal} \sim 1500$~km\,s$^{-1}$, the ICM velocity vector $\vec{v_{\rm ICM}}$ must
have dominant components in the plane of the sky. In the following we give an example:

We define the signs of velocity vectors as positive to the west, north, and away from the
observer. Assuming a galaxy velocity vector of 
$\vec{v_{\rm gal}}=$(-675, -675, 1155)~km\,s$^{-1}$
and taking the total velocity vector from our simulation 
$\vec{v_{\rm tot}}=\vec{v_{\rm gal}}-\vec{v_{\rm ICM}}$=(-2560, -2090, 1155)~km\,s$^{-1}$
leads to a galaxy velocity of $v_{\rm gal}=1500$~km\,s$^{-1}$ and $v_{\rm tot}=3500$~km\,s$^{-1}$
(see Sect.~\ref{sec:bestfit}). The resulting intracluster medium velocity 
is then $\vec{v_{\rm ICM}}=\vec{v_{\rm gal}}-\vec{v_{\rm tot}}$=(1885, 1415, 0)~km\,s$^{-1}$
and the total ICM velocity is $v_{\rm ICM}=2360$~km\,s$^{-1}$.
Thus the motion of the intracluster medium which might be partly due to the infall of the
M49 subgroup is towards the north west and exclusively in the plane of the sky. 
This ICM velocity is 1000~km\,s$^{-1}$ higher than that derived from
X-ray data assuming pressure equilibrium (Shibata et al. 2001).
Alternatively, a local ICM density enhancement of a factor $\sim$4 together with an
ICM velocity of $\sim$1000~km\,s$^{-1}$ can account for the observed stripping radius of NGC~4522.
This is consistent with the radial velocity of M49 ($\sim 1000$~km\,s$^{-1}$) which is very close 
to the velocity of the Virgo cluster. The derived direction of this putative ICM flow might also
be related to the X-ray detection of a gas compression region in the 
north of M49 (Irwin \& Sarazin 1996, Biller et al. 2005)

\section{Conclusions \label{sec:conclusions}}

We present a dynamical model for the evolution of the interstellar medium of
the Virgo cluster spiral galaxy NGC~4522, which presently undergoes strong ram pressure
stripping (Kenney et al. 2004).
We confront the observed with the simulated moment maps. In addition, we solve the
induction equation on the velocity fields of the dynamical model to calculate
the large-scale magnetic field. Assuming a smooth relativistic electron distribution
we obtain the model distribution of polarized radio continuum emission which can be
directly compared to observations (Vollmer et al. 2004).

We conclude that
\begin{itemize}
\item
the model successfully reproduces the observed H{\sc i} gas distribution
and velocity field;
\item
the model fails in reproducing the observed large H{\sc i} linewidths
($\sim 100$~km\,s$^{-1}$);
\item
the MHD model successfully reproduces the observed polarized radio continuum 
emission;
\end{itemize} 

We suggest that large observed H{\sc i} linewidths is due to the warm diffuse
component of the atomic gas which is either stripped more efficiently
or whose velocity dispersion is increased by ram pressure induced shocks.

The model confirms that NGC~4522 undergoes ram pressure which is close to
its maximum. This maximum ram pressure of $\sim 2000$~cm$^{-3}$(km\,s$^{-1}$)$^{2}$
occurred about $50$~Myr ago.
The inclination angle between the galaxy's disk and the ICM wind direction is 
$i=60^{\circ} \pm 10^{\circ}$ and the ICM wind blows from the south-east.
The projected wind angle with respect to the galaxy's minor axis is $6^{\circ} \pm 20^{\circ}$.
Since the galaxy is located at a projected distance of $\sim 1$~Mpc from the 
cluster center where the intracluster medium density is by far too low to cause
significant stripping if it is static, NGC~4522 is either on an unbound orbit within the cluster 
and just passed the region of highest Virgo intracluster medium density or, if the intracluster
medium is moving, the galaxy just leaves the region of highest intracluster
medium velocities. If the intracluster medium is static and smooth the galaxy has a total velocity
of about 3500~km\,s$^{-1}$ with respect to the Virgo cluster. 
If the intracluster medium is moving possibly due to the infall
of the M49 group of galaxies, we derive a velocity of the infalling intracluster medium of
about 2400~km\,s$^{-1}$ with respect to the
Virgo cluster. In this case the intracluster medium associated with the M49 group
is moving towards the north east with a negligible radial velocity component.

This study shows the strength of combining high resolution H{\sc i} and polarized radio continuum
emission with detailed numerical modeling of the evolution of the gas and the large-scale
magnetic field.

\begin{acknowledgements}
This work was supported by Polish-French (ASTRO-LEA-PF)cooperation program,
and by Polish Ministry of Sciences grant PB 378/P03/28/2005.
\end{acknowledgements}

\end{document}